# WEB BASED E-LEARNING IN INDIA: THE CUMULATIVE VIEWS OF DIFFERENT ASPECTS

PARTHA PRATIM RAY[*]

*Department of Electronics and Communication Engineering,
Haldia Institute of Technology, Haldia, Purba Medinipur-721657, West Bengal, India*
*parthapratimray@hotmail.com*

**Abstract**

In the presence of great social diversity in India, it is difficult to change the social background of students, parents and their economical conditions. Therefore the only option left for us is to provide uniform or standardize teaching learning resources or methods. For high quality education throughout India there must be some nation-wide network, which provides equal quality education to all students, including the student from the rural areas and villages. The one and only simple solution to this is Web Based e-Learning. In this paper we try to give some innovative ideas to spread the Web Based e-Learning (WBeL) concept in to the minds of young India along with various approaches taken or to be taken, associated to it till date besides of instructional design models, different course developmental models, the role of technical writing and merit-demerit of WBeL till date.

*Keywords*: Web based learning; India; instructional design model.

## 1. Introduction

India is a country of millions of youth minds, seeking knowledge to move ahead in contrary to their limit. This time is important for India to get prepare for the future hence requiring education in full fledge. Though, we have many schools, enough teachers and facilities for students and teachers. But the great variation in the quality of education is found due to some factors like social background of students, parents, different standards of teacher training programs, all teachers cannot deliver the same message to all learners. This fetches the need of WBeL—Web Based e-Learning.

The term Web Based e-Learning (WBeL) is proposed in this paper because it means its inner meaning,W forWeb orWorld WideWeb—which is an easy media to convey data world-wide very smoothly, B for Based upon W media, e for electronic systems embedded with the W, while L for Learn some data or facts available in the W media Based electronic systems.

As we know that Internet is the ocean of knowledge, therefore it is better to open (introduce) this ocean to all students as early as possible in their life. This can be done by introducing or using Information Technology (IT) and related tools in school education or by using World Wide Web (WWW) as education delivery medium. The WWW is used not only to disseminate information but it also provides a great opportunity to extend learning outside space and time boundaries.

The Web Based Education / Learning has the potential to meet the perceived need for flexible pace, place and face. The web allows education to go to the learner rather than the learner to their education. As per as India is concerned there are many problems that one will face to use IT in education like funds, infrastructure etc.

WWW is a global network and gives the concept of global classroom where any number of students can interact with each other at any time. Goodbye classes, goodbye books and goodbye teachers are all possible with the WBeL. WBeL is an interactive experience with access to on line tutors and can be done from any computers once you have your password. WBeL has become popular amongst educationists because of its inherent strengths and advantages it provides to the instructional process such as, the ability to have multimedia documents, the hypertext/hypermedia capability, WWW network basis, allowing for distance learning.

Access is through web browsers such as Internet explorer and Netscape Navigator. With Web Based Learning, training is organized in the form of modules. The modules are approximately one hour session that focuses on specific subject of training. Using WBeL the training can be brought right to your desktop. This makes technical training more convenient. During the live WBeL module, participants will have the ability to ask the instructor questions, get answers and interact with other students — all on line.

In this paper, we have discussed various innovative methods and challenges according to WBeL in India. In what follows, section 2 presents Indian approaches taken or to be taken on behalf of WBeL. Section 3 presents two new models faced by WBeL to be implied in India. Section 4 presents various instructional design models. Section 5 tells about online learning course development models. Section 6 gives the idea of the impact of technical writing in WBeL. Section 7 presents different advantages and disadvantages of WBeL. Section 8 concludes this paper.

---

[*] Netaji Road, By lane – 2, Newtown, Post + District- Cooch Behar, Pin- 736101, West Bengal, India.

**2. Approaches taken and to be taken**

India has a strong educational infrastructure, particularly in the higher education sector with more than 13, 500 colleges and above 250 universities [11].To cope up with these huge increasing number of education systems we require a strong WBeL system indeed. This section presents the scenario behind this system as already existing and for the near future thoughts has and will be implied on India, respectively.

**2.1.** *Present scenario*

In this portion of paper we present some approaches already taken in account of establishment of WBeL in India so far.

(1) EDUSAT:-

From the use of satellite in the early 1970s to the present interest in a dedicated SATellite for EDUcation (EDUSAT) India has considered education as a primary force for development of the nation. This project has got tremendous success in India during last few years to motivate Indian conventional education to a new era of Hi-Tech education.

(2) *Netvarsity*:-

The countrys first online educational enterprise also came with the private initiative, when the National Institute of Information Technology (NIIT Limited) started Netvarsity in 1996.

(3) Teaching shoppe:-

After what NIIT started in 90s some other private farms came in to the market with a new concept to open up Teaching shoppe (TShopee) for the benefit in the field of school level education and for preparing students for competitive examinations like the medical and engineering entrance tests.

(4) NTFITSD:-

After what NIIT started in 90s some other private farms came in to the market with a new concept to open up the real impetus for e-learning came from the National Task Force on Information Technology and Software Development (NTFITSD) constituted by the Prime Minister of India in 1998. The Task Force report presents the master plan that India has in place as a long term policy for capacity building of institutions, human resource development in IT related areas, and use of ICTs in education.

(5) VCI:-

The Indira Gandhi National Open University (IGNOU) responded to the recommendations of the Task Force with its Virtual Campus Initiatives (VCI) in 1999. Since then a number of such initiatives (Table-1) are in operation in the country.

(6) EEDP:-

 At the Yashwantrao Chavan Maharastra Open University (YCMOU) e-learning is used as a learner support mechanism especially for its Electronics Engineering Diploma Programme (EEDP). Students use a discussion forum to discuss concepts and clarify doubts.

(7) PDF:-

In 2001, the School of Social Sciences at the Indira Gandhi National Open University (IGNOU) started a Post Graduate Certificate in Participatory Management of Displacement, Resettlement and Rehabilitation with the support of the World Bank as a fully online programme that included both synchronous and asynchronous learning opportunities. It is a first of its kind programme, where Participation in Discussion Forum (PDF) is used as a peer evaluation mechanism.

(8) eC-eL:-

Some Other programme features include web course units with interactive exercises, on line Computer Marked Assignments (OCMA), online diary submissions (ODS), e-counselling (Chat) (eC) and e-library (eL) have been introduced in India.

| Name and URL | Areas Covered | Owned and managed by |
|---|---|---|
| Netvarsity http://www.netvarsity.com | IT related areas and soft skills | NIIT Online Learning Limited |
| Indira Gandhi National Open University http://www.ignou.ac.in | IT related areas and in Social Sciences | Part of the National Open University established by the GOI |
| Yashwantrao Chavan Maharastra Open University http://www.ycmou.com | Use e-learning as part of its distance learning strategy in technology courses | Part of YCMOU established by the Maharastra State Government |
| Tamil Virtual University http://www.tamilvu.org | Tamil language, literature, and culture | Governed by the society established by the Government of Tamil Nadu |
| Punjab Technical University http://www.ptuonline.com | Engineering and Technology related courses | Established by the Govt of Punjab, the online venture is a collaborative effort with a trust |
| Birla Institute of Technology and Sciences http://vu.bits-pilani.ac.in | Engineering courses | Part of the BITS (Deemed University status accorded by the University grants Commission) |
| Institute of Management Technology http://www.imtonline.org | Courses in Management leading to eMBA | Managed by NPO, and the courses approved by AICTE |
| Symbiosis Centre for Distance Learning http://www.scdl.net | Courses in Management leading to PG Diploma | Managed by Symbiosis Society (NPO) and approved by AICTE |
| MedVarsity http://www.medvarsity.com | Continuing education programmes in Medical and health related topics | Managed by Medvarsity Online Limited, an unit of Apollo Hospitals |
| Indian Institute of Technology, Mumbai http://www.dep.iitb.ac.in | IT related courses at diploma and noncredit level | Institute of national importance, and the programmes are offered by the Kanwal Rekhi School of Information Technology |
| Indian Institute of Technology, Delhi http://www.iitd.ac.in/courses | Engineering courses | Ministry of Information Technology, Govt. of India supported courses |

Fig. 1. Virtual educational institutions in India [9].

(9) OCPFS:-

Other online initiatives of IGNOU includes online certificate programmes on Food Safety (OCPFS) in collaboration with the Ministry of Health and Family Welfare, Government of India [1], a web-enhanced training package on the Windows version of the UNESCOs popular database management package CDS/ISIS [7], and the Web-based Training Programme for the min career diplomat of the Government of India [1].

(10) NPTEL:-

Other Arguably, the most talked about Indian e-Learning project is the NPTEL project. NPTEL (National Programme on Technology Enhanced Learning) was conceived in 1999 and funded by MHRD (Ministry of Human Resource and Development). Under the project, 7 IITs (Indian Institutes of Technology) and IISc (Indian Institute of Science) Bangalore, worked on the Rs 20.5 crore project from 2003 to 2006, to create 112 video courses and 116 web courses. All these courses are on undergraduate engineering topics, and made to meet most of the requirements of an engineering undergraduate program (at any Indian university). These courses are available to students, working professionals and colleges (both government-aided and private) at virtually no cost or very low cost [10].

Partha Pratim Ray / Indian Journal of Computer Science and Engineering
Vol. 1 No. 4 340-352
(11)  *Amrita Vishwa Vidyapeetham*-- This initiative launched in 2004 uses satellite technology to connect 4 campuses of Amrita University located in 4 cities of South India. There is collaboration with US universities also, and the project was expected to expand to 200 universities. It was based on technological support from ISRO [3].

(12)  *BITS Pilani*-- It has established a virtual university, with DIT sponsorship. BITS has been one of the pioneers in distance education. BITS has been providing courses for working professionals in distance education mode leveraging technology [3].

(13)  *Jadavpur University*-- It started a new inter-disciplinary Masters in Multimedia Development course in 2000-01, as a distance education course using print material, CD ROM, and web-based learning environment. Technology was provided by CDAC Kolkata and CMC.

(14)  *Aligarh Muslim University*-- It worked on a project in 2006-07 to take its distance education program online, starting with a few courses which are industry-relevant.

(15)  *Central Institute of English and Foreign Language*, Hyderabad It had a project for online learning software set-up and usage in 2006.

(16)  Another commercially successful initiative is MBA Programs being conducted for Working Professionals using Satellite Video technology, by institutions like IIM-Calcutta, IIM-Calicut, IIT-Delhi, IIFT, IIT Bombay, XLRI etc. This was done by these institutions using services provided by companies like HughesNet (formerly Hughes Direcway), Reliance Infocom and now NIIT Imperia.

(17) Many other universities and colleges had had small projects/ initiatives where they bought software/hardware and other technology products, got content development done for e-Learning launch. It included the likes of Hyderabad University, Kerala University, Terna College Mumbai, MDI Gurgaon, etc.

(18)  In India, the University of Madras opened a Virtual University in partnership with University of Mumbai and University of Calcutta. This Virtual University system has led to the commencement of 10 joint degrees, post graduate and Ph.D programmes [14].

**2.2. *Other concepts that can be applied to India***

(1)  *ICOL:-*
Establishment of Indian Council for on line Learning [9] in recent year as a statutory body to.

- Develop a national distributed repository of Reusable Learning Objects (RLO).
- Maintain standard for on line Learning.
- Coordinate and promote on line Learning.
- Accreiate On line Learning courses and programmes.
- Develop and maintain on line learning portals for life-long learning.

(2)  *eLC:-*

Establishment of e-Learning Consortium (eLC) including member educational institutions to offer e-learning programmes without duplicating efforts to spread the ideas of e-Learning among Indians.

(3)  *OTT:-*

On line Training Teachers (OTT) is a programme that will help Indian teachers, diverse by geography, polity, culture and brilliance to unify one in all respect.

*(4)  DSLOT:-*

DSLOT is an approach to facilitate Development of Small re-usable Learning Objects by Teachers, to share the ideas and perceptions behind the learning objects through a sharable web portal.

ISSN : 0976-5166    343

(5) *WBCMS:-*

The urgency to create courses in response to the growing demand for online learning has resulted in a hurried push to drop PowerPoint notes into Web based course management systems (WBCMSs) [19], devise an electronic quiz, put together a few discussion questions, and call it a course. Web-based learning environments can serve as motivational, instructional, modeling, feedback, and assessment tools.

(6) *KMWBLE:-*

Knowledge mining is the discovery of hidden knowledge stored possibly in various forms and places in large data repositories. Knowledge is a valuable asset to most organizations as a substantial source to enhance organizational competency [13]. A top-level of knowledge management model in web-based learning environment is shown in Figure 2.

(7) *SCROM:-*

Shareable Content Object Reference Model [15] is currently adopted by a lot of industrial and educational organizations as a set of standards to specify course structure and content delivery process. SCORM is a collection of standards that enable interoperability, accessibility and reusability of distributed heterogeneous Web based learning systems. Reusability in SCORM model can be achieved through the concept of learning objects.

(8) *KMWBLE:-*

One popular application has been for educational use, such as Web-based Distance Distributed or online Learning ($WD^2L$). The use of the Web as an educational tool has provided learners and educators with a wider range of new and interesting learning experiences and teaching environments, not possible in traditional in class education [8]. Figure 3 shows features of $WD^2L$.

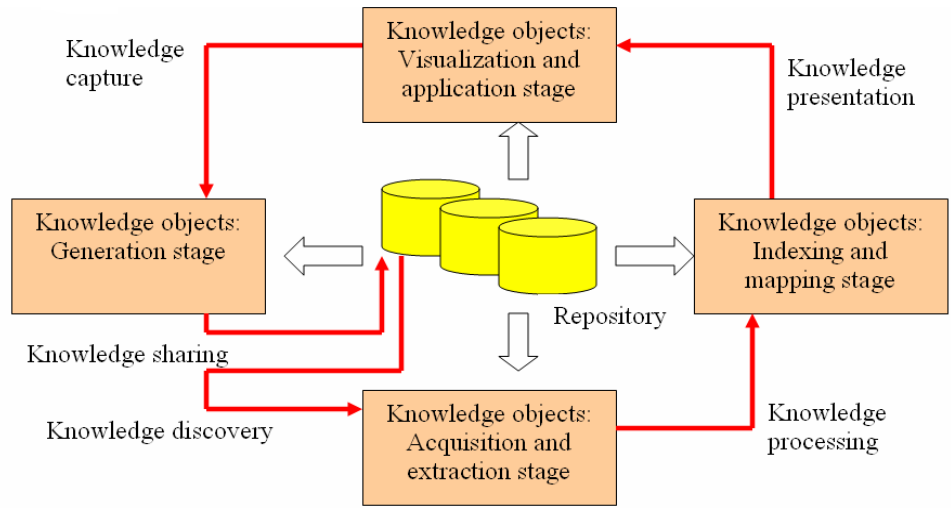

Fig. 2.  A top-level management model of knowledge objects in Web Based Learning.

(9)  We should open up new universities like British Open University, with its headquarters in Milton Keynes, UK was started in 1971.With an initial enrollment of 25,00 students it immediately became the second largest in UK, next only to University of London [4]. Today, British Open University has more than 178,000 students and is the largest in UK. More importantly, for the third year in a row, it had the highest satisfaction scores from students (95%, as measured in September 2007) [5]. Further 50,000 organizations have sponsored staff for Open University courses.

| Feature | Relationship to WD²L Environment | Component |
|---|---|---|
| Interactive | • Allow interactions with students, instructors, and Web resources via various communication channels<br>• Provide interactive feedback on students' performance | Discussion Board<br>Practice Sessions<br>Quiz |
| Multimedia | • Support students' various learning styles using a variety of multimedia | Concept Map<br>Text to Speech<br>Advanced Organizers |
| Distributed | • Allow downloading and printing the materials from the WD²L environment and any other Web sources | GPS Resources<br>GPS Glossary |
| Collaborative Learning | • Create a medium of collaboration, conversation, discussion, exchange, and communication of ideas | Discussion Board (By Group) |

Fig. 3.  Features of WD²L.

### 3. New models

Till date various types of systems including electronic and theoretical have been proposed and implied but each has its own demerits. We try to give some innovative ideas that we can apply on India to make it WBeL-India.

#### 3.1. *CSIES*

Central State Interactive Education System (CSIES) is a new idea that we are going to elaborate and employ in this paper. As per our design shown in Figure 4, we have divided the CSIES in four segments as below.

(1) *Central Server*:-

The top most part of CSIES is crowned by N number of *Central Servers* or CS, through whole India based on different mega cities and capitals of states. Each CS will be interconnected such type of more N-1 number of servers, which will be governed by central education ministry of government of India. The Central Servers will be connected to N number of State Servers through Intranet , named CSIES-Intra-I.

(2) *Local Server:-*

Each *Local Server* (LS) will be connected internally to each other. Their access privilege will be given to different state ministry of education respectively. Each LS will have four sub-divisions having different grade of education connected by Intranet named CSIES-Intra-II.

(3) *Information Provider Center:-*

*Information provider Center* (IPC) will be the key among all hierarchy of CSIES in India. IPC will get divided into four units namely, primary, high, graduation, post graduation. Each of these will get connected to IPC by means of Internet, firewall and CSIES-Intra-III.

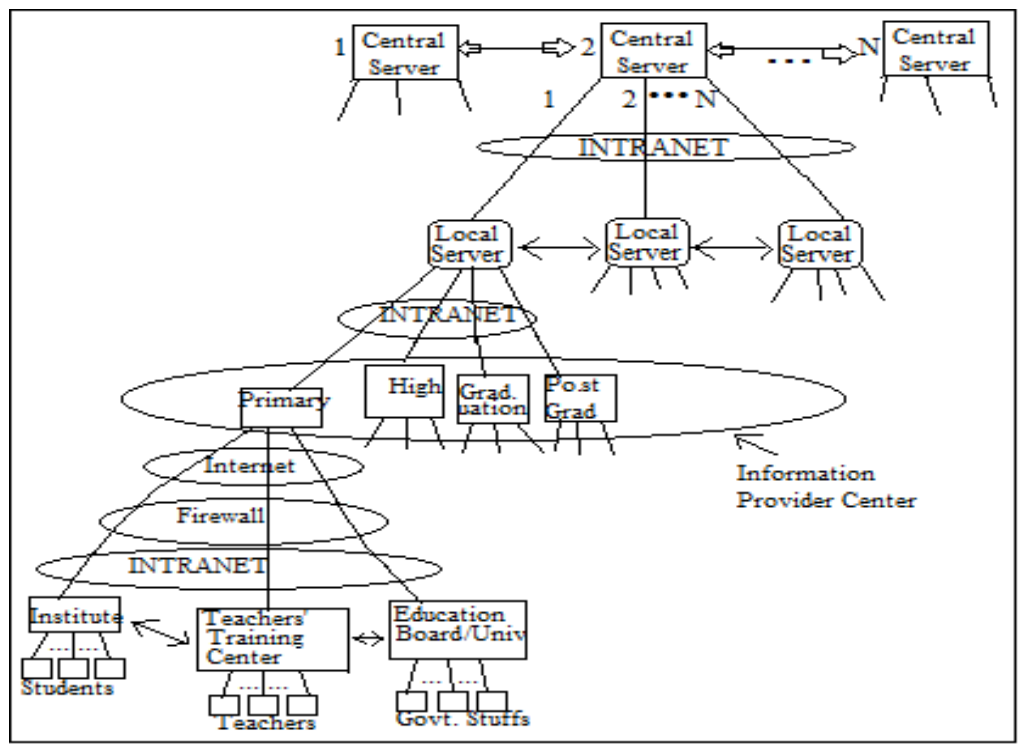

Fig. 4.   Central State Interactive Education System  (CSIES).

(3)  *End User Access:-*

The lower most stair of CSIES hierarchy is conserved by End User Access (EUA), basically a group of servers, which will be of three types as below.

- Institute: - This server provides all information such as course, syllabus, examination schedule, results, interactive learning and all other education related real facts to the students as a school, college, or university.
- Teachers Training Center:- This server provides all information regarding training of teachers education and workshop that will help Indian teachers to bring under one umbrella synchronization and motivation towards a same goal, so that cultural, geographical, and linguistic mismatch can be easily overridden.
- Education Board/ University: - This kind of server will be directly connected to the government stuffs so that they can look after the whole CSIES and education system.

**3.2.  *MBES***

The mobile revolution is finally here in the form of Mobile Based Education System (MBES) a mobile learning approaches an extension to WBeL in India. In the figure below the logical architecture of MBES is shown. We can divide the whole MBES into three parts as server, central system, end-user. The fundamental concept behind the MBES is illustrated below in figure 5.

(1)  *Server:-*

This portion of MBES serves the requests to the central part regarding various aspects such as connectivity between end user and control system, validation and verification of end user, authorization, accessibility of end user through a well established Intranet that connects all the machines of this portion. Regardless of the Intranet there will be Internet connectivity between the server and central system of MBES.

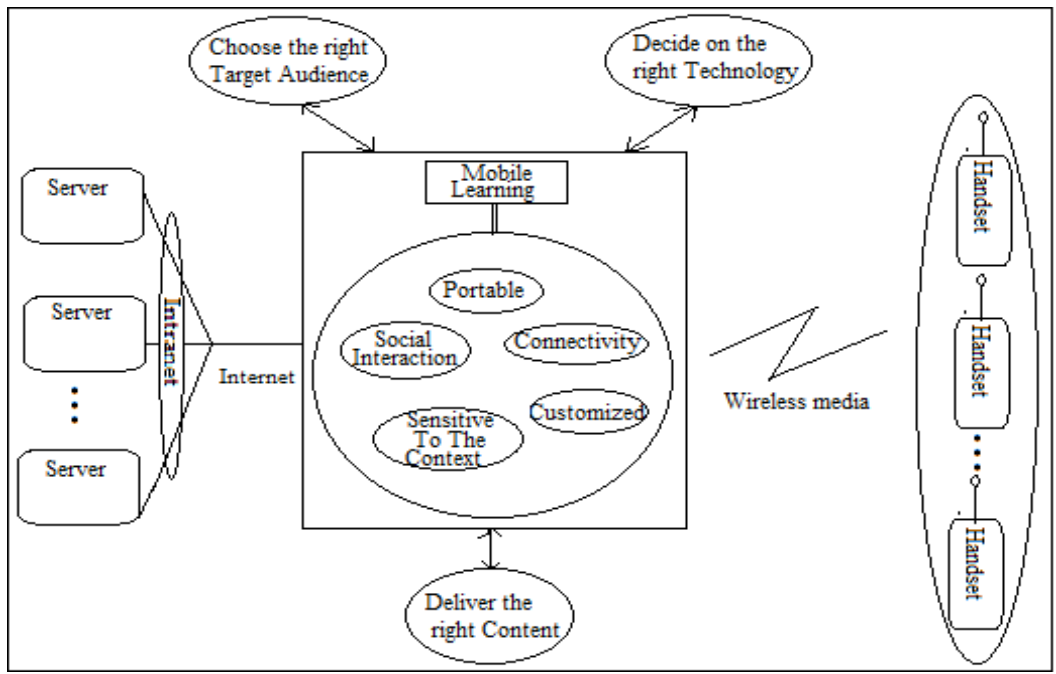

Fig. 5. The architecture of Mobile Based Education System.

(2) *Central system*:-

There are five basic parameters for production and development of mobile based central system as portable the ease access of information through a mobile phone or a PDA, connectivity is backbone of this system helping the connection between the strong network and hand held devices, social interaction exchange of data with other people( friends, colleagues etc.)and gain considerable knowledge, customized to make useful of customized learning information and sensitive to the context gathers real and simulated data unique to the current location, environment, and time. The three important tasks that the central system must go mostly in order to perform its job well are as choose the right target audience, deliver the right content, and decide on the right technology.

(3) *End user*:-

In MBES the last portion is the end user where the mobile and other hand held devices (PDA, net-book etc.) having multimedia and internet connectivity to the central system through wireless media based upon modern technologies such as GPRS, Broadband etc.

**4. Instructional design models for WBeL**

At the root of Instructional Design and/or Instructional Design Models, is a systematic process that Instructional Designers should follow in order to achieve the creation of efficient and effective instruction. Or more simply put, Instructional Design (ID) is a framework for learning [16]. This framework asks the Instructional Designer to assess the desired outcomes of the learning and begin to apply an ID model that is most appropriate to assist in achievement of these desired outcomes. Despite some ID models being quite generic in nature, they are incredibly popular and capable because they present a very effective, yet general, model to build various types of instruction to meet different objectives in learning.

Below we will see a variety of popular models listed. These items do not attempt to outline the specifics of any Instructional Design model, but rather serve to convey the variety and possible application of these models to your specific instructional task. As you may notice, or soon come to learn, most of these models can be modified to meet your specific needs. Their systematic frameworks allow you to borrow from their strengths and retrofit several models to meet your differing needs.

(1) *ADDIE (Assess Design Develop Implement Evaluate):-*
- Very generic, yet very successful.
- Probably one the most followed models.

(2) *Algo-Heuristic:-*
- This theory suggests all cognitive activities can be analyzed into operations of an algorithmic (measure of complexity), semi-algorithmic, heuristic (computational method), or semi-heuristic nature.
- Once these operations are determined, they can form the basis of instructional strategies and methods.
- Don't just teach knowledge, but the algorithms and heuristics of experts as well.

(3) *Dick and Carey Model:-*
- Breaks instruction down into smaller components.
- Used to teach skills and knowledge.

(4) *Robert Gagńe's ID Model:--*
- Gagńe's Nine Events of Instruction.
    a) Gain attention
    b) Inform learners of objectives
    c) Stimulate recall of prior learning
    d) Present the content
    e) Provide *learning guidance*
    f) Elicit performance (practice)
    g) Provide feedback
    h) Assess performance
    i) Enhance retention and transfer to the job

(5) *Minimalism:-*
- Developed by J.M. Carroll.
- Framework to design instruction specific to computer users.
- Learning tasks should be meaningful and selfcontained activities.
- Learners should be given realistic projects.
- Instruction should permit self-directed reasoning and improvising.
- Training materials and activities should provide for error recognition and recovery.
- Provide a close linkage between the training and actual system.

(6) *Kemp, Morrison, and Ross:-*
- Nine step instructional design model
    a) Identify instructional problems
    b) Examine learner characteristics
    c) Identify subject content
    d) State instructional objectives
    e) Sequence content within each instructional unit for logical learning
    f) Design instructional strategies
    g) Plan the instructional message and delivery
    h) Develop evaluation instruments to assess objectives
    i) elect resources to support instruction and learning activities

(7) *Rapid Prototyping (Rapid E-Learning):-*
- Learners and/or subject matter experts interact with prototypes and instructional designers in a continuous review and revision process.
- Development of a prototype is the first step.
- Analysis is continuous throughout the process.

(8) *Empathic Instructional Design:-*
- 5-step process [16]
    a) Observe

b) Capture data
c) Reflect and analyze
d) Brainstorm for solutions
e) Develop prototypes

## 5. On line learning course development models

The choice of a particular approach to the development of an online-learning course is based on several factors including the academic tradition and resources available to the organization. Institutions that are dedicated to online and distance education have tended to adopt a more collaborative course team approach. Conventional campus-based educational providers, on the other hand have tended to adopt a lesser collaborative approach. In any event, the development of an online-learning course comprises a new experience for many. It calls for new skills such as in e- moderation and some de-skilling as well (i.e., shedding off of old lecturing habits). Old habits die hard, and when faced with circumstances that render some of ones previous experience irrelevant there is quite a lot of uneasiness, loss of confidence, disillusionment, hostility, and at times ithdrawal from the activity altogether.

### 5.1. *Wrap around model*

This model of online-learning relies on study materials, which may comprise online study guides, activities and discussion wrapped around existing previously published resources such as textbooks or CD-ROMs etc. This model represents a resource-based approach to learning, as it seeks to use existing material that is relatively unchanging and is already available online of offline. Such courses, once they are developed, can be taught or tutored by persons other than the course developers. Collaborative learning activities in the form of group work, discussion among peers, and online assessments may be supported by computer conferencing, or mailing lists [11][12].Unfortunately quite often, these online learning elements tend to be added to the course and do not form an integral part of the assessment requirements of the course.

### 5.2. *The integrated model*

This model is closest to a full online-learning course. Such courses are often offered via a comprehensive learning management system. They comprise availability of much of the subject matter in electronic format, opportunities for computer conferencing, small group-based collaborative online learning activities, and online assessment of learning outcomes. For the moment though, some of the subject matter content will be best-accessed offline in already published textbooks and other sources. The learning and teaching in these courses takes place in the computer conferences, in which the prescribed readings and the assigned tasks are discussed. Much of this learning and teaching activity is fairly fluid and dynamic as it is largely determined by individual and group activities in the course. To some extent, this integrated model dissolves the distinctions between teaching and learning in favor of the facilitation of learning [2].

## 6. The role of technical writing in e-learning

Many companies today are looking for the most cost-effective way to train their employees. By utilizing e-learning, companies save money by lessening employee travel expenses and limiting employee time away from work. Computer-Based Training (CBT) and Web-Based Training (WBT) are two solutions. CBT training traditionally involves use of a CDROM; in many cases online help is directly available with this approach. WBT is training delivered via the Internet.

With the development of e-learning technologies technical writers have become more in demand. They have an increasing role in the design, development and implementation of training. It is vital for a technical writer to have good writing skills, but equally important, they must have the ability to produce, test, and implement their materials using sophisticated software.

### 6.1. *First Steps*

A very important first step for a successful technical writer is to work in collaboration with management and any involved individuals or departments to gather information. This needs to occur before writing one sentence! The technical writer must be provided the mission/goals for the project, receive input for a course outline, and understand who the target audience will be. Once this has occurred, the writer will lay out a course outline, storyboards, and scripting, select learning activities, and produce media when indicated. The last step involves testing, evaluating, and finalizing the material.

### 6.2. *Development Tools*

The writer will use special tools for development, including:

- Instructional design software which lays out instruction design principles. An example of this software is Designer's Edge, a popular training design and planning tool.
- Course management and testing tools which manages a course and provides controlled tests. An example of this software is ASPTESTS.
- Web page tools to design web pages and websites. Examples include Dreamweaver, along with a course build-in for Dreamweaver, FrontPage, and Flash, which allows the creation of quickly downloaded animation.
- Multimedia applications to enhance your Web pages. Examples include, but are not limited to, Adobe Illustrator, Corel Draw, Adobe Photoshop, and Paint Shop Pro (by JASC), one of the original graphics editing software.

### 6.3. *Writing Skills*

Good writing skills cannot be minimized! It is important to use sound guidelines and common sense:

- For excellent organization, plan sections and subsections well.
- Use good layout for easy reading such as color fonts for main headings.
- Avoid background and history information.
- Use the first page to present the most important information.
- Word headings and subheadings with strong verbs and nouns. These command attention and tell the reader exactly what you're covering in any given section.
- Aim for a simple approach. Use plain English and simple words. Make your document concise and easy to read.
- Use active verbs versus passive verbs. Active verbs make a document shorter, simpler to read, and easier to comprehend.
- Avoid jargon and technical terms.
- Acronyms and abbreviations become annoying when readers aren't aware what they mean.
- Use only well known abbreviations such as IBM or Washington, DC.
- Keep sentence length short; 10 to 20 words.
- Break down longer sentences in list form for readability.
- Avoid wordy phrases; make every word count.
- Use plenty of examples and illustrations; a picture is worth a thousand words.

### 6.4. *Evaluate and Test*

Evaluate and test your material when it is complete! Ask individuals from your target audience to test your material by following the steps outlined. Many times this results in rethinking, redesigning, and possibly rewriting sections of your material.

## 7. Advantages and disadvantages of WBeL

### 7.1. *Advantages*

E-learning is beneficial to education, corporations and to all types of learners. It is affordable, saves time, and produces measurable results. E-learning is more cost effective than traditional learning because less time and money is spent traveling. Since e-learning can be done in any geographic location and there are no travel expenses, this type of learning is much less costly than doing learning at a traditional institute.

Flexibility is a major benefit of e-learning. E-learning has the advantage of taking class anytime anywhere. E education available when and where it is needed. E-learning can be done at the office, at home, on the road, 24 hours a day, and seven days a week. . E- learning also has measurable assessments which can be created so the both the instructors and students will know what the students have learned, when they've completed courses, and how they have performed.

Students like e-learning because it accommodates different types of learning styles. Students have the advantage of learning at their own pace. Students can also learn through a variety of activities that apply to many different learning styles learners have. Learners can fit e-learning into their busy schedule. If they hold a job, they can still be working with e-learning. If the learner needs to do the learning at night, then this option is available. Learners can sit in their home in their pajamas and do the learning if they desire.

E-learning encourages students to peruse through information by using hyperlinks and sites on the worldwide Web. Students are able to find information relevant to their personal situations and interest. E-learning allows students to select learning materials that meet their level of knowledge, interest and what they need to know to perform more effectively in an activity. E-learning is more focused on the learner and it is more interesting for the learner because it is information that they want to learn. E-learning is flexible and can be customized to meet the individual needs of the learners.

E-learning helps students develop knowledge of the Internet. This knowledge will help learners throughout their careers. E-learning encourages students to take personal responsibility for their own learning. When learners succeed, it builds self knowledge and self-confidence in them.

Educators and corporations really benefit from e-learning. Learners enjoy having the opportunity to learn at their own pace, on their own time, and have it less costly.

### 7.2. *Disadvantages*

Next we look at the disadvantages of e-learning. One disadvantage of e-learning is that learners need to have access to a computer as well as the Internet. They also need to have computer skills with programs such as word processing, Internet browsers, and e-mail. Without these skills and software it is not possible for the student to succeed in e- learning. E-learners need to be very comfortable using a computer. Slow Internet connections or older computers may make accessing course materials difficult. This may cause the learners to get frustrated and give up. Another disadvantage of e-learning is managing computer files and online learning software. For learners with beginner-level computer skills it can sometimes seem complex to keep their computer files organized. Without good computer organizational skills learners may lose or misplace reports causing them to be late in submitting assignments. Some of the students also may have trouble installing software that is required for the class.

E-learning also requires just as much time for attending class and completing assignments as any traditional classroom course. This means that students have to be highly motivated and responsible because all the work they do is on their own. Learners with low motivation or bad study habits may fall behind. Another disadvantage of e-learning is that without the routine structures of a traditional class, students may get lost or confused about course activities and deadlines causing the student to fail or do poorly.

Another disadvantage of e-learning is that students may feel isolated from the instructor. Instructions are not always available to help the learner so learners need to have discipline to work independently without the instructor's assistance. E-learners also need to have good writing and communication skills. When instructors and other learners aren't meeting face to face it is possible to misinterpret what was meant.

## 8. Conclusion

In this paper we tried to give some information about different approaches towards WBeL in India and two different new models CSIES and MBES to be implied on Indian education system to make it more web based hence advanced. We discussed other topics such as instructional design models, online learning course development, and impact of technical writing along with various merits and demerits of WBeL. Recently we are seeking out other aspects of WBeL and their relation to Indian society.